# Simulating Organogenesis in COMSOL: Parameter Optimization for PDE-based models

Denis Menshykau[1,2], Srivathsan Adivarahan[1], Philipp Germann[1,2], Lisa Lermuzeaux[1,3], Dagmar Iber[*1,2]
[1]D-BSSE, ETH Zurich, Switzerland;
[2]Swiss Institute of Bioinformatics (SIB), Switzerland;
[3]Department of Bioengineering, University of Nice-Sophia Antipolis, France;
*Corresponding author: D-BSSE, ETH Zurich, Mattenstrasse 26, 4058, Switzerland, dagmar.iber@bsse.ethz.ch

**Abstract:** Morphogenesis is a tightly regulated process that has been studied for decades. We are developing data-based and image-basd mechanistic models for a range of developmental processes with a view to integrate the available knowledge and to better understand the underlying regulatory logic. In our previous papers on simulating organogenesis with COMSOL (German et al COMSOL Conf Proceedings 2011; Menshykau and Iber, COMSOL Conf Proceedings 2012) we discussed methods to efficiently solve such models on static and growing domains. A further challenge in modeling morphogenesis is the parameterization of such models. Here we discuss COMSOL-based methods for parameter optimization. These routines can be used to determine parameter sets, for which the simulations reproduce experimental data and constraints. Such data is often image based, but may also come from classical biochemical or genetic experiments.

## 1. Introduction

Increasing computational power and advancement of computational methods now allow us to formulate and solve detailed computational models for gene regulatory networks that control complex biological processes in morphogenesis and organogenesis. The reliability of developmental processes suggests that core processes are deterministic, and accordingly deterministic models for pattern formation have been studied for decades in developmental biology.

We formulate the models as systems of reaction-diffusion equations of the form:

$$\dot{X_i} + \nabla \cdot (u \cdot X_i) = D_i \nabla^2 X_i + R_i$$

where $u$ denotes the velocity field of the domain and $R_i$ the reactions, which couple the equations for the different species $X_i$. $D_i$ is the diffusion constant and $\nabla$ the Nabla operator. The velocity field might be either imposed, or could be made dependent on the local concentration of proteins, which may affect local cell behavior, e.g. the cell division rate or cell adhesion and motility. A wide range of reaction laws can be used.[1] We have previously shown how to implement and to efficiently solve such models in COMSOL on growing and static multilayer domains.[2,3] The computational models can help to integrate the available knowledge, to test the consistency of current models, and to generate new hypotheses. Previously, we correctly predicted novel genetic regulatory interactions in the limb bud,[4] suggested mechanisms for the control of digit formation,[5] bone development,[6] and for branch point selection in the lung[7] and kidney.[8]

One major challenge is the parameterization of computational models.[9] Parameters that are required for the formulation of models include, but are not limited to protein production and degradation rates, diffusion constants, and rate constants for protein-protein interactions. These parameter values can, in principle, be measured *in vitro* and sometimes also *in vivo*. However, this is rarely the case and the establishment of measurement methods for a certain system can take years and can typically not be easily transferred to another system. Therefore, direct measurement of all required parameter values is not feasible.

Experiments in developmental biology typically deliver information about the organ shape and protein expression patterns at various stages in a WT and mutant system. Here we explore the possibility to use these data to quantitatively test, parameterize and discretize computational models. First we consider a simple Turing type model formulated on an idealized 2D domain. Next we discuss examples from image based modeling of branching morphogenesis.

## 2. Conventions & Computational Details

Here we use **bold italic** to refer to COMSOL fields and nodes e.g. *Coefficient form PDE*, *Diffusion Coefficient* refers to the field where the diffusion coefficient needs to be specified. Models discussed in this manuscript were implemented through the *Coefficient form PDE* interface and solved in COMSOL 4.3a and 4.3b as previously described in ref.[3]

## 3. Use of COMSOL Multiphysics
### 3.1 A Test Case – Simple Turing Type Model

We have previously proposed that Turing type models[10,11] based on receptor-ligand signaling govern lung and kidney branching morphogenesis[7,8]. The simplest form of such a model for the receptor ligand-interactions can be represented by Schnakenberg-type reaction kinetics:

$$\dot{U} = \Delta U + \gamma(a - U + U^2V)$$
$$\dot{V} = D\Delta + \gamma(b - U^2V) \quad (1)$$

where *a, b, γ, D* are positive parameters and *D>>1*.

*V* describes the distribution of ligand and is defined in both layers (Figure 1a); however the ligand production rate is zero in the epithelium (*b*=0). *U* describes the distribution of receptor and is defined only in the epithelium (inner layer in Figure 1b). $U^2V$ represents the concentration of the ligand-receptor complex (Figure 1c).

Turing type models generate patterns spontaneously from homogenous, noisy initial conditions; the solution of eq 1 is depicted in Figure 1. Depending on the parameter values different patterns emerge (compare the left and right columns in Figure 1).

To test the optimization procedure we choose to optimize the distribution of the ligand-receptor complex, $U^2V$, along the epithelium-mesenchyme border (Figure 1c). We first construct the cost function as $L^2$ distance between the calculated distribution $U^2V$ for a given parameter set and $U_0^2V_0$ calculated earlier at *a*=0.1, *b*=0.4 and *γ*=300.

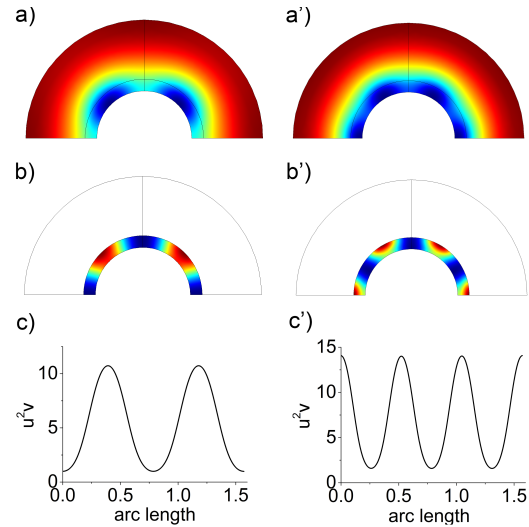

**Figure 1. Test Case: a Turing type model on a two-layer domain.** Steady state distribution of a) the variable V (ligand) and b) the variables *U* (receptor). c) Distribution of $U^2V$ along the border of the epithelium and mesenchyme. *D*=100, *a*=0.3, *b*=0.5. Panels with and without apostrophe where calculated for γ=300 and γ=500 accordingly.

$$\Delta = \sqrt{\int_L (U_0^2V_0 - U^2V)^2 dl} \quad (2)$$

$U_0^2V_0$ was specified using *Functions-> Interpolation:* fc(x), where fc(x) is the tabulated solution $U_0^2V_0$. The deviation $\Delta^2$ was specified as *Boundary Probe, type Integral*: (u1^2*v1-fc(x))^2.

We first tested the gradient-based optimization solver SNOPT:[12,13] (*Optimization-> Method: SNOPT*). *Optimality tolerance* was set to $10^{-3}$; *Optimization Solver->Gradient Method* was set to *Numeric*. The other settings were left at their default values. Next we randomly sampled 100 points, which were set as initial value for the optimization solver (*Optimization->Control Variable and Parameters-> Initial Value*). Points were sampled from a log uniform distribution in the range log(*a)*: -2..0, log(b*)*: -1.4..0.6 and log(γ): 1.4..3.4. To automatically run the optimization algorithm starting from various initial values we ran COMSOL via LiveLink for MATLAB.

Initial values of parameters were set in COMSOL with the function *param.set()*; values of parameters after optimization where retrieved from COMSOL with function *mphglobal()*. The distribution of variables $U$ and $V$ where retrieved from COMSOL with the function *mphevelal()* and the number of peaks with *findpeaks()*. Next, we tested the gradient free optimization method **Coordinate Search: (Optimization-> Method: Coordinate Search)**.[14] In this case Initial values of parameters were set with the function *model.study('std1').feature('opt').set('initval', values_str)*.

Figure 2 a, c) shows that the optimization solver SNOPT correctly recovers the distribution of $U^2V$ only in a confined region of the parameter space, around $\gamma$=300. Figure 3a depicts the value of the cost function, $\Delta$, in the parameter space.

We notice that the parameter region, where a minimal value of $\Delta$ is observed, is similar to the region of the parameter space where the optimization solvers converge (Figure 2). Figure 3b shows the number of $U^2V$ peaks in the Turing pattern for the different parameter values. This region of the parameter space is again similar to that depicted in Figure 2 a. It thus seems that the optimization solver correctly recovers parameter values only if the pattern that is calculated for the initial parameter set has already the correct number of $U^2V$ peaks. However, a more detailed analysis as presented in Table 1, shows that the optimization solver SNOPT can recover the correct value of the parameters also if the initial pattern has two or one peaks or lies outside of the Turing range. We next tested the gradient-free optimization solver Coordinate Search (Figure 2 b, d)) and found that, similar to SNOPT, correct values of the initial parameters can be recovered only from a confined region of the parameter space. However, in the case of the Coordinate Search algorithm the region is confined around $b=0.4$.

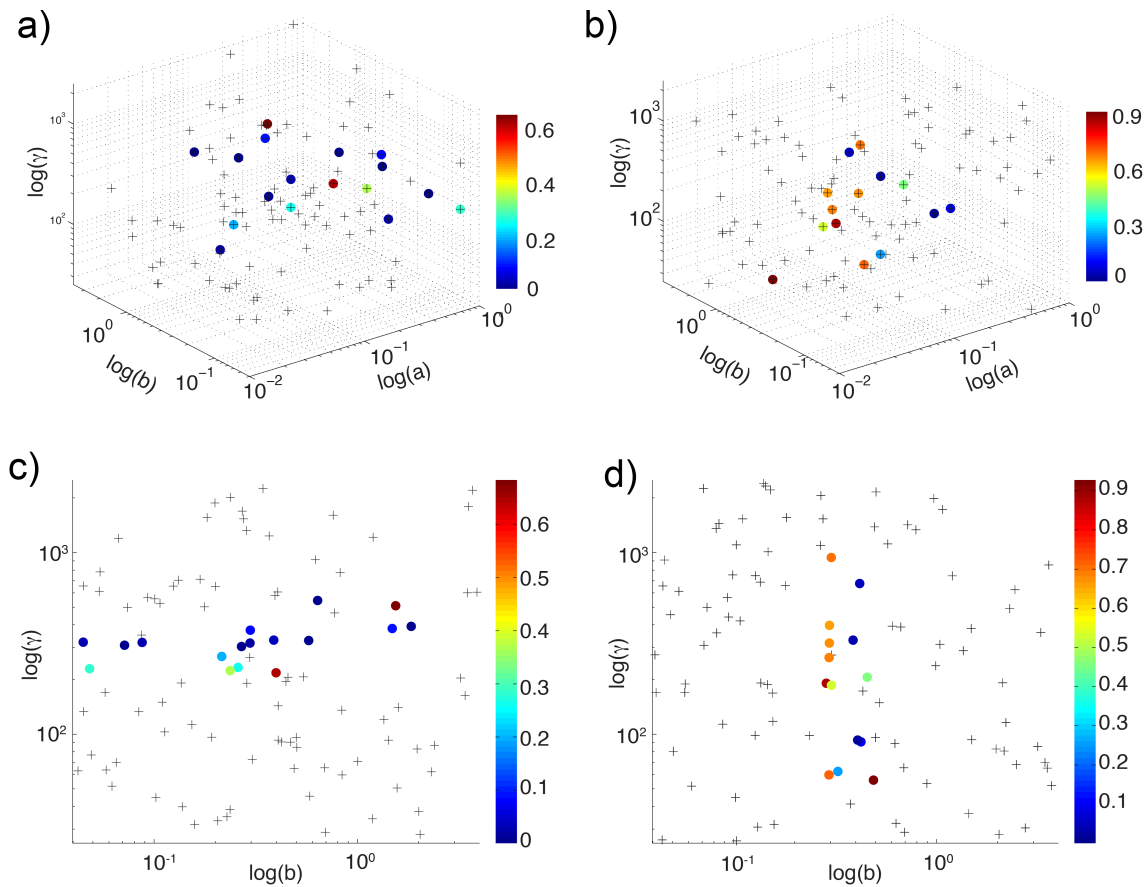

**Figure 2. Convergence of the Optimization Solvers.** Optimization solver a and c) SNOPT, b, d) Coordinate Search. Points and crosses depict initial values for the optimization solver, which lead to the convergence and failure of the optimization solver, accordingly. Color code shows value of the objective function at the end of the optimization. Panels c) and d) depict projection of 3D parametric space on $\gamma$-$b$ plane.

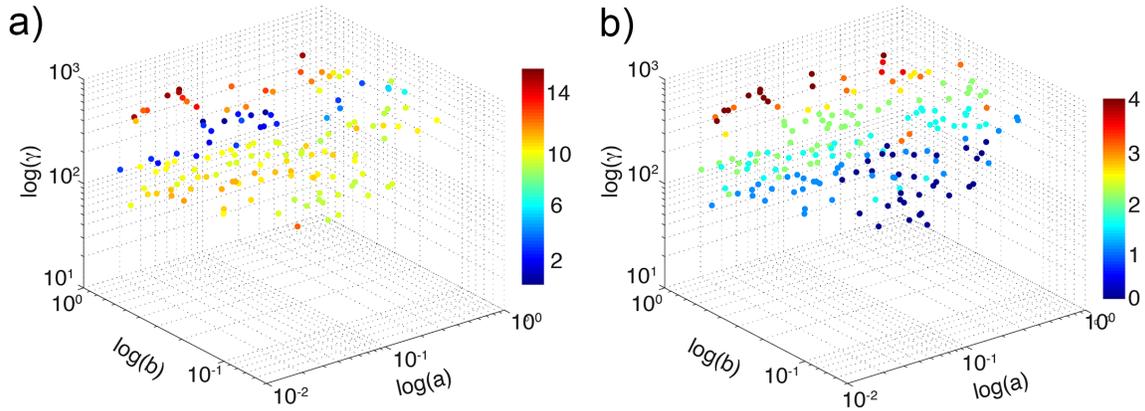

**Figure 3. Landscape of parameter space of the Turing-type model (eq 1).** a) Deviation, $\Delta$ (eq 2) and b) number of peaks in $U^2V$ distribution.

Coordinate Search Method and SNOPT both experience a similar drawback: the parameter values can be recovered only from a confined region of the parameter space. To overcome this drawback we suggest the following strategy for parameter optimization of Turing-type models: 1) uniformly sample the parameter space of interest, 2) choose points with the minimum value of cost function, and 3) use these points as a starting condition for the local optimization solver e.g. SNOPT or Coordinate Search.

### 3.2 Application: Kidney Branching Morphogenesis

A number of alternative mechanisms have been proposed to govern branching morphogenesis.[15] However, to date no consistent test of alternative models has been carried out. Here we deploy an image-based modeling approach[16,17] to allow the testing of such models.

**Table 1. Convergence of the Optimization Solver SNOPT.** Initial ($a_0$, $b_0$, $\gamma_0$) and optimized parameter values ($a_{opt}, b_{opt}, \gamma_{opt}$). Parameters used to generate data for fit are $a=0.1, b=0.4, \gamma=300$. $n_0$ refers to the number of peaks in the U2V distribution.

| $a_0$ | $b_0$ | $\gamma_0$ | $n_0$ | $\Delta_0$ | $a_{opt}$ | $b_{opt}$ | $\gamma_{opt}$ | $n_{opt}$ | $\Delta_{opt}$ |
|---|---|---|---|---|---|---|---|---|---|
| 0.068 | 0.297 | 317.4 | 2 | 2.122 | 0.123 | 0.393 | 317.4 | 2 | 3E-4 |
| 0.219 | 1.481 | 381.7 | 0 | 46.21 | 0.188 | 0.370 | 381.7 | 2 | 0.091 |
| 0.131 | 0.388 | 329.4 | 2 | 0.051 | 0.138 | 0.389 | 329.4 | 2 | 0.032 |
| 0.566 | 0.072 | 308.5 | 0 | 7.588 | 0.100 | 0.400 | 299.9 | 2 | 8E-5 |
| 0.236 | 1.539 | 509.1 | 0 | 67.28 | 0.205 | 0.331 | 509.2 | 2 | 0.235 |
| 0.467 | 0.576 | 327.8 | 0 | 6.404 | 0.136 | 0.389 | 327.8 | 2 | 0.002 |
| 0.314 | 0.398 | 217.7 | 1.5 | 7.444 | 0.010 | 0.454 | 217.7 | 2 | 0.482 |
| 0.656 | 0.298 | 373.6 | 2 | 4.000 | 0.099 | 0.400 | 299.9 | 2 | 8E-5 |
| 0.027 | 0.215 | 268.0 | 0 | 8.092 | 0.100 | 0.407 | 300.3 | 2 | 4E-4 |
| 0.096 | 0.259 | 232.9 | 0 | 8.040 | 0.025 | 0.433 | 232.9 | 2 | 0.257 |
| 0.067 | 0.636 | 544.8 | 0 | 11.89 | 0.099 | 0.400 | 299.6 | 2 | 0.002 |
| 0.412 | 0.236 | 224.0 | 1 | 8.362 | 0.029 | 0.438 | 226.7 | 2 | 0.369 |
| 0.062 | 1.837 | 392.0 | 0 | 76.78 | 0.099 | 0.400 | 299.7 | 2 | 2E-4 |
| 0.177 | 0.045 | 321.0 | 0 | 7.578 | 0.128 | 0.392 | 321.0 | 2 | 0.023 |
| 0.619 | 0.269 | 303.4 | 1.5 | 6.784 | 0.105 | 0.400 | 303.6 | 2 | 0.005 |
| 0.793 | 0.048 | 229.3 | 0 | 7.750 | 0.020 | 0.433 | 230.2 | 2 | 0.281 |
| 0.010 | 0.087 | 320.3 | 0 | 320.4 | 0.123 | 0.392 | 320.4 | 2 | 0.023 |

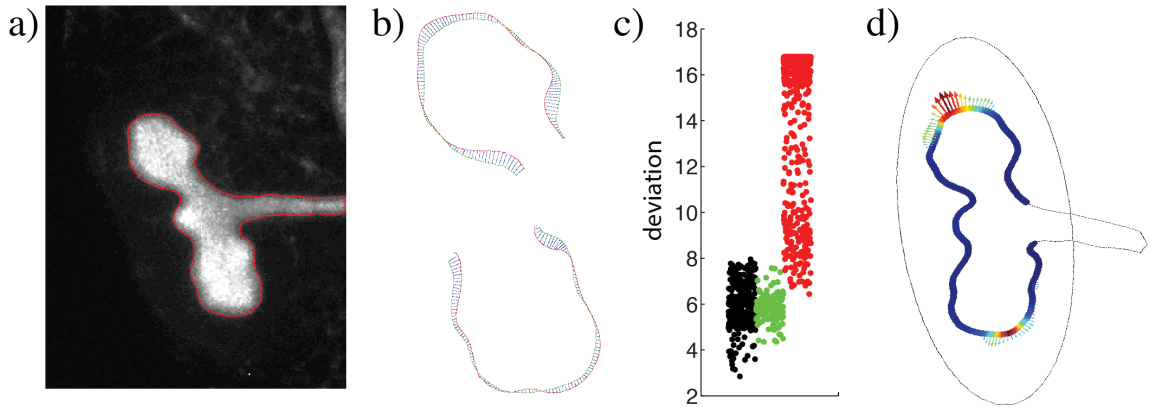

**Figure 4. Image-based Modelling of Kidney Branching Morphogenesis**. a) A snapshot from the time lapse movie; red line indicates the extracted border of the kidney epithelium. Experimental data is courtesy of Odysse Michos; b) enlarged parts of the kidney explant and the calculated displacement field (blue), green and red lines shows earlier and later frames in the time lapse movie, accordingly; c) deviation (eq 2) for the points in the Turing space (black), in between (green), and out of the Turing space (red); d) distribution of $U^2V$ on the epithelium-mesenchyme border shown by the color code (blue - low concentration, red - high concentration), arrows indicate growth field.

We proposed a mechanism for branch point selection during lung[7] and kidney[8] branching morphogenesis. The mechanism is based on receptor-ligand interactions, which lead to a Turing type instability and domain patterning. To test whether this mechanism can correctly predict growth areas during branching morphogenesis we obtained 2D time lapse movies of the embryonic kidneys *in vitro* and extracted domain boundary in Matlab (Figure 4a).[17] We also calculated displacement fields between consecutive stages (Figure 4b). The displacement field represents the position-dependent direction and magnitude of the tissue growth.[16,17] Various algorithms can be applied to calculate displacement fields, e.g. landmark-free algorithms based on minimal or normal distances, or the landmark-based Buckstein algorithm.[18] Figure 4b depicts the displacement field calculated with an algorithm based on the normal distance between surfaces.

To build a computational model we imported the domain in the shape of the embryonic kidney explants (***Geometry -> Interpolation Curve-> Import to Table***) and the displacement field into COMSOL (***Global -> Function- > Interpolation -> File***). The displacement field was next used in COMSOL to specify the cost function in the form analogous to that given by eq 2. To estimate parameter values we used the approach outlined in section 3.1 A Test Case – Simple Turing Type Model: first we sampled a thousand points (Figure 4c) from a log normal distribution, and we next performed a local optimization using points in the parameter space with the minimum value of the cost function as a starting point. The signalling model (eq 1) formulated on the two-layer domain yielded a better fit for parameter values inside the Turing space (Figure 4c, black dots) than outside the Turing space (Figure 4c, red dots).[15,17] Figure 4d shows that the Turing type model adequately reproduces the growth areas observed in the embryonic kidneys. Similarly 3D imaging data can be used to solve, parameterize and discretize models.[16]

## 4. Discussion

Here we presented an approach for model parameterization based on a random sampling of the parameter space followed, by a local optimization.

The discussed models were solved on a static 2D domain, which allows solving them fast. In particular, the solution time for the model depicted in Figure 4d is several seconds; the solution time for a similar 3D image-based model is several minutes.[16] However, organogenesis is a highly dynamic process, involving dramatic changes in the shape of organs. Therefore, models should be formulated on a growing and deforming domain.

Preliminary tests show that the solution of such models requires hours of computational time.[17] Therefore, extensive parameter screening becomes prohibitive. The development of more efficient and well-parallelized algorithms for both model parameterization and solution on a deforming domain is required for further advancement of computational organogenesis.

## 5. References

1. Iber, D. and Fengos, G. Predictive models for cellular signaling networks in Computational Modeling of Signaling Networks (Editor Xuedong Liu), Humana Press.
2. Germann P., Menshykau D., Tanaka S. and Iber D. Simulating Organogenesis in COMSOL, *Proc COMSOL Users Conf Stuttgart* (2011).
3. Menshykau D. and Iber D. Simulating Organogenesis in COMSOL: Deforming and Interacting Domains, *Proc COMSOL Users Conf Milan* (2012).
4. Probst, S., Kraemer, C., Demougin *et al* SHH propagates distal limb bud development by enhancing CYP26B1-mediated retinoic acid clearance via AER-FGF signalling, *Development,* 138, 1913-23 (2011)
5. Badugu A, Kraemer C, Germann P, Menshykau D and Iber D Digit patterning during limb development as a result of the BMP-receptor interaction. *Scientific Reports*, 2:991
6. Tanaka S and Iber D, Inter-dependent tissue growth and Turing patterning in a model for long bone development *Phys Biol, in press*
7. Menshykau D., Kraemer C. and Iber D. Branch Mode Selection during early Lung Development *PLoS Comp Biology*, 8(2): e1002377 (2012).
8. Menshykau D. & Iber D. Kidney branching morphogenesis under the control of a ligand-receptor based Turing mechanism *Phys. Biol. 10 (2013) 046003.*
9. Geier F, Fengos G, Felizzi F and Iber D Analysing and constraining signaling networks: parameter estimation for the user Springer Book Series: Methods in Molecular Biology. Book Title: Computational Modeling of Signaling Networks (Editor Xuedong Liu)
10. Turing, A. M. The Chemical Basis of Morphogenesis. *Phil Trans Royal Soc London* 237 (641): 37–72 (1952).
11. Murray J. D. *Mathematical Biology II*. Springer.
12. P.E. Gill, W. Murray, and M.A. Saunders, User's Guide for SNOPT Version 7: Software for Large-Scale Nonlinear Programming, Systems Optimization Laboratory (SOL), Stanford University, 2006.
13. P.E. Gill, W. Murray, and M.A. Saunders, SNOPT: An SQP Algorithm for Large-Scale Constrained Optimization, SIAM Review, vol. 47, no. 1, pp. 99–131, 2005.
14. A.R. Conn, K. Scheinberg, and L.N. Vicente, Introduction to Derivative-Free Optimization, MPS-SIAM Series on Optimization, SIAM, 2009.
15. Iber D and Menshykau D, The Control of Branching Morphogenesis *Open Biol* 2013 3:130088.
16. D. Iber, S. Tanaka, P. Fried, P. Germann, D. Menshykau "Simulating Tissue Morphogenesis and Signaling" Springer Book Series: Methods in Molecular Biology; Book Title: Tissue Morphogenesis: Methods and Protocols (Editor Celeste Nelson)
17. Adivarahan S, Menshykau D, Michos O and Iber D Dynamic Image-Based Modelling of Kidney Branching Morphogenesis". Lecture Notes in Computer Science (Springer), Computational Methods in Systems Biology (CMSB) 2013 IST Austria.
18. F. L. Bookstein, Principal warps: Thin-plate splines and the decomposition of deformations, Pattern Analysis and Machine Intelligence, 1989.

## 9. Acknowledgements

We would like to thank the COMSOL support team and especially Sven Friedel, Thierry Luthy, Karthik Reddy, Reine Granström, Mahesh Kadam, Jonas Nilsson, Anders Ekerot, Hanna Gothäll, Linus Andersson and Reine Granström for their support and insightful comments; Rolf Zeller and his group (and in particular Erkan Ünal) for the experimental data and the ongoing collaboration; Odysse Michos for the experimental data and the ongoing collaboration. The authors acknowledge funding from SNF Sinergia grant "Developmental engineering of endochondral ossification from mesenchymal stem cells".